\documentclass[hidelinks,a4paper,11pt]{article}
\usepackage[utf8]{inputenc}

\usepackage{authblk}

\usepackage{hyperref}
\hypersetup{colorlinks,
            pdftitle={On The Evolution Of User Support Topics in Computational Science and Engineering Software},
            pdfauthor={K.~Rupp, S.~Balay, J.~Brown, M.~Knepley, L.~C.~McInnes, B.~Smith},
            pdfsubject={Computational Science & Engineering Software Sustainability and Productivity Challenges} }

\usepackage{fullpage}
\usepackage{graphicx}

\title{\vspace*{-2.8cm}On The Evolution Of User Support Topics \\ in Computational Science and Engineering Software}
\author{\vspace*{-.35cm}K.~Rupp$^1$, S.~Balay$^2$, J.~Brown$^{2,3}$, M.~Knepley$^{4,5}$, L.~C.~McInnes$^2$, and B.~Smith$^2$}
\date{\vspace*{-.45cm}$^1$ Institute for Microelectronics, TU Wien, Austria \\
      $^2$ Mathematics and Computer Science Division, Argonne National Laboratory, USA \\
      $^3$ Laboratory for Advanced Numerical Simulation, University of Colorado Boulder, USA \\
      $^4$ Computation Institute, University of Chicago, USA \\
      $^5$ Computational and Applied Mathematics, Rice University, USA \\ \vspace*{-.65cm}}

\begin{document}

\maketitle

\begin{abstract}
We investigate ten years of user support emails in the large-scale solver library PETSc in order to identify changes in user requests.
For this purpose we assign each email thread to one or several categories describing the type of support request.
We find that despite several changes in hardware architecture as well programming models, the relative share of emails for the individual categories does not show a notable change over time.
This is particularly remarkable as the total communication volume has increased four-fold in the considered time frame, indicating a considerable growth of the user base.
Our data also demonstrates that user support cannot be substituted with what is often referred to as \emph{better documentation} and that the involvement of core developers in user support is essential.
\end{abstract}

\textbf{Introduction.}
User support and community interaction is a crucial ingredient for the success software packages in computational science and engineering (CSE)~\cite{Bangerth:SuccessfulCSESoftware,Turk:Scaling-Human-Dimension}.
With several disruptive changes to computing such as the paradigm shift to multi-core processors or the use of graphics processing units for general purpose computations in recent years, one may expect a substantial change in the type of user support requests for software libraries tailored to the efficient use of computing resources in CSE.
Similarly, one may expect that the type of user requests changes as the user base of a software package grows.

In the following we provide answers to these questions for the large-scale solver library PETSc\footnote{PETSc webpage: \url{http://www.mcs.anl.gov/petsc/}} by systematically evaluating email archives for the years 2006 to 2014.
Our study is based on the public \emph{petsc-users} mailinglist\footnote{Archive available online: \url{http://lists.mcs.anl.gov/pipermail/petsc-users/}}, which is the primary list for general user support.

\textbf{Email Categorization.}
All email discussions for the years 2006 to 2014 have been systematically assigned to one or more of the following categories:
\begin{itemize}
 \setlength\itemsep{0.2em}
 \item \textbf{Algorithms}:
   In-depth discussions and general advice on the best mathematical approaches for setting up a solver for a given problem.

 \item \textbf{Beginner}:
   Short questions related to documented functionality or on how to deal with basic operations for distributed parallel systems in general.

 \item \textbf{Bug}:
   Bug reports by users on either the release version of PETSc or new functionality which has not been officially released yet.
   
 \item \textbf{Features}:
   Discussions about whether PETSc provides a certain feature, about future directions, or about better explanation of exising features from a user's perspective.
   
 \item \textbf{Performance}:
   Email communication about potential performance enhancements for a given machine or hardware architecture.
   In contrast to the \emph{Algorithmic} section, the user is primarily interested in reducing execution times for an existing solver setup rather than the design of new types of solvers for the problem at hand.
   
 \item \textbf{Runtime Errors}:
   Discussions triggered by runtime errors.
   These may be due to errors in user code such as memory corruption, invalid use of functionality provided by PETSc, or missing functionality in PETSc.
\end{itemize}
We do not include configuration and installation issues here, because these are directed to the \emph{petsc-maint} mailing list.
Since the numbers of emails on \emph{petsc-users} and \emph{petsc-maint} are comparable, configuration and installation issues would be the dominating category if included in the list above.
However, full data for the considered timeframe for \emph{petsc-maint} is not available, yet empirical checks showed that the same trends hold true.
For the sake of completeness we note that developer discussions on the \emph{petsc-dev} mailinglist are about half of the email volume on \emph{petsc-users} and \emph{petsc-maint}.

\begin{figure}
 \centering
 \includegraphics[width=0.75\textwidth]{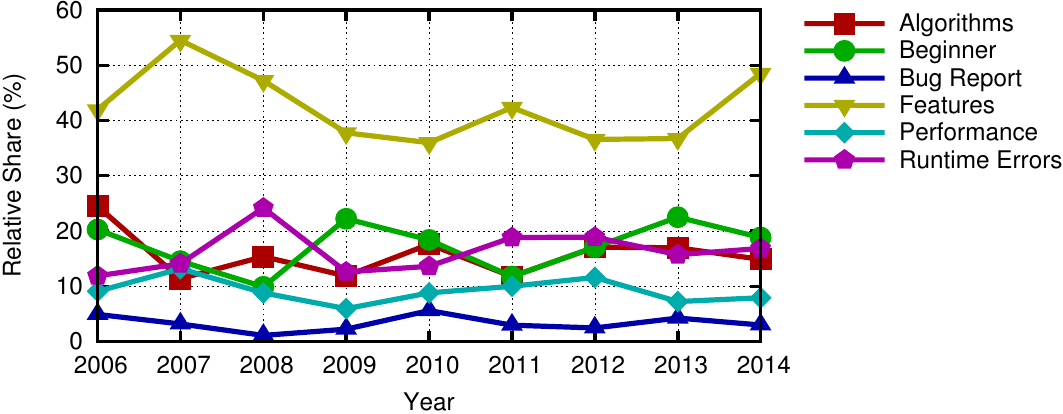}
 \caption{Categorization of email threads on the \emph{petsc-users} mailing list.}
 \label{fig:emails}
\end{figure}

\textbf{Results and Discussion.} 
The change in the relative share of email threads assigned to each category is depicted in Figure~\ref{fig:emails}.
Even though there is an inherent fluctuation in the relative share, none of the categories shows a significant net increase or reduction over the full time frame considered.
Questions on features represent the largest category and amount to about 40 percent of the total email threads.
The nature of these feature questions is such that these questions cannot be answered entirely through improved documentation.
Thus, even if PETSc developers devoted all their time to documentation only, the total amount of support emails would not be reduced significantly.

Considering that the total number of emails to \emph{petsc-users} was 964 in 2006 and 3947 in 2014, it is remarkable that the relative share of each category has changed by at most eight percent for each category when comparing 2006 with 2014.
This suggests that the user base of PETSc has grown uniformly without preference for either highly experienced or less experienced users.

Overall, the results of our evaluations clearly underline that user support is an often underestimated ingredient for successful CSE software.
Maintaining a sustainable software stack for CSE therefore not only requires funding for implementing new features from scratch, but also needs to account for the efforts required for preserving the health and integrity of existing packages.
Also, our data shows that it is essential that core developers are involved in user support:
All email threads in the category \emph{Algorithms} and most in the category \emph{Features} require in-depth knowledge of the library, hence core developers are essential for at least half of all email discussions.

\bibliographystyle{plain} 
\bibliography{ref}

\begin{thebibliography}{1}

\bibitem{Bangerth:SuccessfulCSESoftware}
W.~Bangerth and T.~Heister.
\newblock {What Makes Computational Open Source Software Libraries Successful?}
\newblock {\em Computational Science \& Discovery}, 6:015010/1--18, 2013.

\bibitem{Turk:Scaling-Human-Dimension}
M.~J. Turk.
\newblock {Scaling a Code in the Human Dimension}.
\newblock In {\em Proceedings of the Conference on Extreme Science and
  Engineering Discovery Environment: Gateway to Discovery}, XSEDE '13, pages
  69:1--69:7, New York, NY, USA, 2013. ACM.

\end{thebibliography}

\end{document}